\documentclass[lettersize,journal]{IEEEtran}
\usepackage{amsmath,amsfonts}
\usepackage{algorithmic}
\usepackage{array}
\usepackage{textcomp}
\usepackage{stfloats}
\usepackage{url}
\usepackage{verbatim}
\usepackage{graphicx}
\usepackage{cite}
\usepackage{amsthm}
\usepackage{tabularx}
\usepackage{multirow}
\usepackage{amssymb}

\usepackage{hyperref}
\def\BibTeX{{\rm B\kern-.05em{\sc i\kern-.025em b}\kern-.08em
    T\kern-.1667em\lower.7ex\hbox{E}\kern-.125emX}}
\usepackage{balance}
\begin{document}
\title{Rate-Splitting Multiple Access with a SIC-Free Receiver: An Experimental Study}
\author{Guoqian Sun, Xinze Lyu, Bruno Clerckx%
\thanks{G. Sun, X. Lyu, and B. Clerckx are with the Department of Electrical and Electronic Engineering, Imperial College London, London SW7 2AZ, U.K. (e-mail: \{g.sun24, x.lyu21, b.clerckx\}@imperial.ac.uk).}}
\maketitle
\bstctlcite{IEEEexample:BSTcontrol}
\markboth{IEEE Communications Letters}{Sun \MakeLowercase{\textit{et al.}}: Rate-Splitting Multiple Access with a SIC-Free Receiver: An Experimental Study}
\begin{abstract}
 Most Rate-Splitting Multiple Access (RSMA) implementations rely on successive interference cancellation (SIC) at the receiver, whose performance is inherently limited by error propagation during common-stream decoding. This paper addresses this issue by developing a SIC-free RSMA receiver based on joint demapping (JD), which directly evaluates bit vectors over a composite constellation. Using a two-user Multiple-Input Single-Output (MISO) prototype, we conduct over-the-air measurements to systematically compare SIC and JD-based receivers. The results show that the proposed SIC-free receiver provides stronger reliability and better practicality over a wider operating range, with all observations being consistent with theoretical expectations.
\end{abstract}


\section{Introduction}
\IEEEPARstart{F}{uture} wireless communication systems are required to support high spectral efficiency and massive connectivity under heterogeneous user conditions. In the downlink multi-user multiple-input multiple-output (MU-MIMO) setup, the design of multiple access schemes is crucial for mitigating inter-user interference and improving system robustness. Conventional Spatial Division Multiple Access (SDMA) suffers from performance degradation under spatially correlated channels~\cite{1261332}. Meanwhile, Non-Orthogonal Multiple Access (NOMA) is less effective when users have similar channel strengths~\cite{6692652}, as insufficient power disparity weakens the effectiveness of successive interference cancellation (SIC).

Rate-Splitting Multiple Access (RSMA) has emerged as a promising multiple access framework to overcome these limitations. Rate-splitting was initially studied for single-antenna interference channels~\cite{1056307} and was later extended to multi-antenna broadcast channels, where it was shown to outperform conventional schemes under partial Channel State Information (CSI) at the Transmitter~\cite{Joudeh2016PartialCSIT}. By splitting each user’s message into a common stream decoded by all users and private streams decoded individually, RSMA bridges the gap between fully decoding interference and treating interference as noise~\cite{Mao2018RSMA}. This flexible interference management strategy has been shown to enlarge the achievable rate region compared to SDMA and NOMA, improve spectral efficiency and fairness, and enhance robustness against channel variations and imperfect CSI~\cite{Hao2015FiniteFeedback,Joudeh2016PartialCSIT,9491092}. As a result, RSMA has been extensively investigated in terms of transceiver design and resource optimization under various CSI assumptions~\cite{9831440}.

Despite these theoretical advances, research on practical RSMA receiver designs remains relatively limited. Most existing studies adopt SIC-based receivers and assume Gaussian codebooks, while practical systems operate with finite constellations and Bit-Interleaved Coded Modulation~\cite{10609431}. SIC is sensitive to residual interference and prone to error propagation, where errors in the earlier decoding stage can severely degrade subsequent decoding. Motivated by this gap, recent theoretical work has proposed SIC-free RSMA receiver designs, among which joint demapping (JD) achieves near-optimal performance by directly evaluating bit-level Log-Likelihood Ratios (LLRs) over a composite constellation and decodes the common and private streams without serial interference cancellation \cite{10485496,11351316}. These studies demonstrate that SIC-free receivers, such as JD-based receivers can achieve consistent performance gains over conventional SIC under harsh channel conditions; however, their validation has been limited to theoretical analysis and link-level simulations.

On the other hand, RSMA prototyping research using Software-Defined Radio (SDR) platforms have experimentally demonstrated the advantages of RSMA over SDMA and NOMA in unicast, multicast, and integrated sensing and communication scenarios \cite{10471302,10718324,11358849}. Nevertheless, all existing RSMA SDR prototypes adopt SIC-based receivers, leaving the practical feasibility and performance of SIC-free RSMA receivers unexplored. To the best of the authors’ knowledge, no experimental prototype has yet implemented and evaluated a SIC-free RSMA receiver.

In this letter, we address this open problem by realizing a SIC-free RSMA receiver based on JD on a wideband multi-user MISO SDR prototype. Using over-the-air measurements, we compare SIC and JD-based receivers in terms of sum throughput and coded bit error rate (BER) under different channel conditions and spatial correlations. The main contributions are as follows:

\begin{enumerate}

\item We design and implement a SIC-free RSMA receiver based on joint demapping on a practical wideband SDR platform with finite constellations.

\item We conduct systematic over-the-air measurements to compare JD and SIC-based receivers under various Modulation and Coding Scheme (MCS) settings and channel conditions.

\item Experimental results demonstrate that the proposed JD-based receiver achieves smoother and more reliable sum throughput, preserves private-message decodability when the common stream is under pressure, and reduces the decoding Signal-to-Noise Ratio (SNR) threshold.
\end{enumerate}

\section{JD-based RSMA System Model}
We follow the first SDR-based RSMA prototype architecture in \cite{10471302}. We consider a two-user MISO downlink, where the transmitter sends messages $W_1$ and $W_2$ to user-1 and user-2 respectively. 
At the transmitter, each message $W_k$ with $k\in\{1,2\}$ is first divided by the message splitter into a common component $W_{c,k}$ and a private component $W_{p,k}$. 
The two common components are then combined into a single common message $W_c$. 
After channel coding and modulation, $W_c$ produces a common stream. 
In parallel, each private component $W_{p,k}$ is encoded and modulated on its own, which forms the private stream $s_{p,k}$. 
The three streams, $s_c$, $s_{p,1}$, and $s_{p,2}$, are linearly precoded to generate the OFDM frequency-domain transmit signal:
\begin{equation}
\begin{aligned}
\mathbf{x}[n] &= \mathbf{p}_c s_c[n] + \mathbf{p}_1 s_{p,1}[n] + \mathbf{p}_2 s_{p,2}[n],
\end{aligned}
\label{eq: tx signal}
\end{equation}
where $[n]$ indicates the symbol carried by the $n$-th sub-carrier, $n\in\{0,...,N_c-1\}$, $\mathbf{p}_c$, $\mathbf{p}_1$, and $\mathbf{p}_2$ denote the linear precoders assigned to the common stream and to the two private streams.

On the receiver side, let $\mathcal{X}_c$ denote the modulation set of the common stream and $\mathcal{X}_{p,k}$ the modulation set of user $k$'s private stream, $k\in\{1, 2\}$. After  Cyclic Prefix (CP) removal and Discrete Fourier Transform processing, the received signal at user-$k$ is modeled as:
\begin{equation}
\label{eq:2}
\begin{aligned}
\mathbf{y}_k[n]
&=\mathbf{h}_k^{\!H}[n]\mathbf{x}[n]+n_k[n]\\
&= \mathbf h_k^{H}[n]\mathbf p_c\, s_c[n]
+\mathbf h_k^{H}[n]\mathbf p_1\, s_{p,1}[n]\\
&\quad+\mathbf h_k^{H}[n]\mathbf p_2\, s_{p,2}[n]+n_k[n],
\end{aligned}
\end{equation}
where $n_k[n]$ represents the thermal noise, whose variance is denoted by $\sigma^2$ and in practice it is estimated from pilots and residual error statistics. 

Before joint demapping, we define the scalar effective channel coefficient $g$ on subcarrier $n$ as the projection of the estimated MISO channel onto a transmit precoder, which captures the combined amplitude scaling and phase rotation at the receiver. Accordingly, the effective gains seen by the common and private streams at user $k$ on subcarrier $n$ are:
\begin{equation}
 g_{c,k,n}=\mathbf{h}_k^{H}[n]\mathbf{p}_c,\qquad g_{p,k,n}=\mathbf{h}_k^{H}[n]\mathbf{p}_k.
\end{equation}
For each subcarrier $n$, the receiver utilizes these effective gains to form the composite constellation:
\begin{equation}
\mathcal{S}_{k,n}=\Big\{g_{c,k,n}\,x_c+g_{p,k,n}\,x_{p,k},\;x_c\in\mathcal{X}_c,\ x_{p,k}\in\mathcal{X}_{p,k}\Big\}.
\end{equation}

User-$k$ subsequently uses a joint demapper on each subcarrier $n$ to select a composite symbol pair $(s_c[n],s_{p,k}[n]) \in \mathcal{X}_c \times \mathcal{X}_{p,k}$, which enables receiver $k$ to decode common and private streams simultaneously. Within the SIC-free receiver, each user does not perform successive subtraction of the common stream before decoding its private stream. Instead, the bit vectors of both the common and private streams are obtained directly from the \emph{composite constellation} in a single joint demapping stage, where the contributions of different streams are jointly considered rather than successively canceled. This eliminates the hierarchical decoding dependency inherent to SIC-based receivers.

At user-$k$, the joint demapper outputs two sets of bit vectors that are sent to:
\begin{enumerate}
  \item the channel decoder for $W_c$, producing an estimate of the common message, $\widehat{W}_{c,k}$;
  \item the channel decoder for $W_{p,k}$, producing an estimate of the private message $\widehat{W}_{p,k}$.
\end{enumerate}
In the end, $\widehat{W}_{c,k}$ and $\widehat{W}_{p,k}$ are combined to form user $k$’s estimate of its desired message, $\widehat{W}_k$.

To keep the analysis parallel to the SIC-based prototype, we adopt the same wideband coded-OFDM convention with polar code as in~\cite{10471302}. Specifically, to generate each stream $s\in\{s_c,s_{p,1},s_{p,2}\}$, a single polar codeword is interleaved, modulated and mapped across all $N_c$ OFDM subcarriers within one frame. Under this setting, deep fades on some data-bearing subcarriers may dominate the decoding reliability. As a theoretical baseline for transmitter-side benchmarking, we consider a simple bottleneck proxy to provide a conservative wideband reference for MCS benchmarking. We define the per-stream wideband spectral-efficiency proxy as:
\begin{equation} \label{eq:proxy}
    R_s \triangleq \min_{n \in \{0,\dots,N_c-1\}} \log_2(1 + \text{SINR}_s[n]),
\end{equation}
where $\text{SINR}_s[n]$ is the per-subcarrier Signal-to-Interference-plus-Noise Ratio (SINR). For the common stream, we further use $\text{SINR}_{s_c}[n] = \min_k \text{SINR}_{c,k}[n]$ to reflect that the common message must be decoded by both users.

Importantly, this proxy serves primarily as a theoretical baseline. In practice, our SDR prototype supports a finite set of MCSs, and we evaluate system performance in bits per second (bps) to align with hardware capabilities and realistic deployment metrics. To bridge the discrete physical layer parameters with the continuous system data rate, we consider the system effective bandwidth $B_{eff}$ and an MCS group $\mathcal{M} = \{(m_c, r_c), (m_1, r_1), (m_2, r_2)\}$. Let $R(m,r) \triangleq B_{eff} \cdot m \cdot r$ denote the nominal wideband data rate of a selected stream, where $m = \log_2(M)$ is the number of bits per symbol and $r$ is the code rate. The expected JD sum throughput (in bps) is therefore formulated as:
\begin{equation} \label{eq:throughput}
    \begin{aligned}
        T_{\text{mcs}}^{\text{JD}}(\mathbf{P}, \mathcal{M}) &= R(m_c, r_c) \Pr(\hat{W}_{c,1} = W_c, \hat{W}_{c,2} = W_c) \\
        &\quad + \sum_{k=1}^{2} R(m_k, r_k) \Pr(\hat{W}_{p,k} = W_{p,k}),
    \end{aligned}
\end{equation}
where $\mathbf{P} \triangleq \{\mathbf{p}_c, \mathbf{p}_1, \mathbf{p}_2\}$. The common stream contributes to the throughput only when both users decode it successfully, while each private stream contributes according to the decoding success of its intended user.

Based on the expected sum throughput defined in \eqref{eq:throughput}, we evaluate the system performance by exhaustively scanning a standardized MCS subset. For each MCS selection, the decoding success probabilities are obtained empirically from measurements. This procedure ensures a fair comparison between SIC and JD, with all results grounded in measured decoding outcomes rather than theoretical capacity limits.

\section{JD-based RSMA Prototype Design and Implementation}
\begin{figure}[!t] 
    \centering
    \includegraphics[width=1\columnwidth]{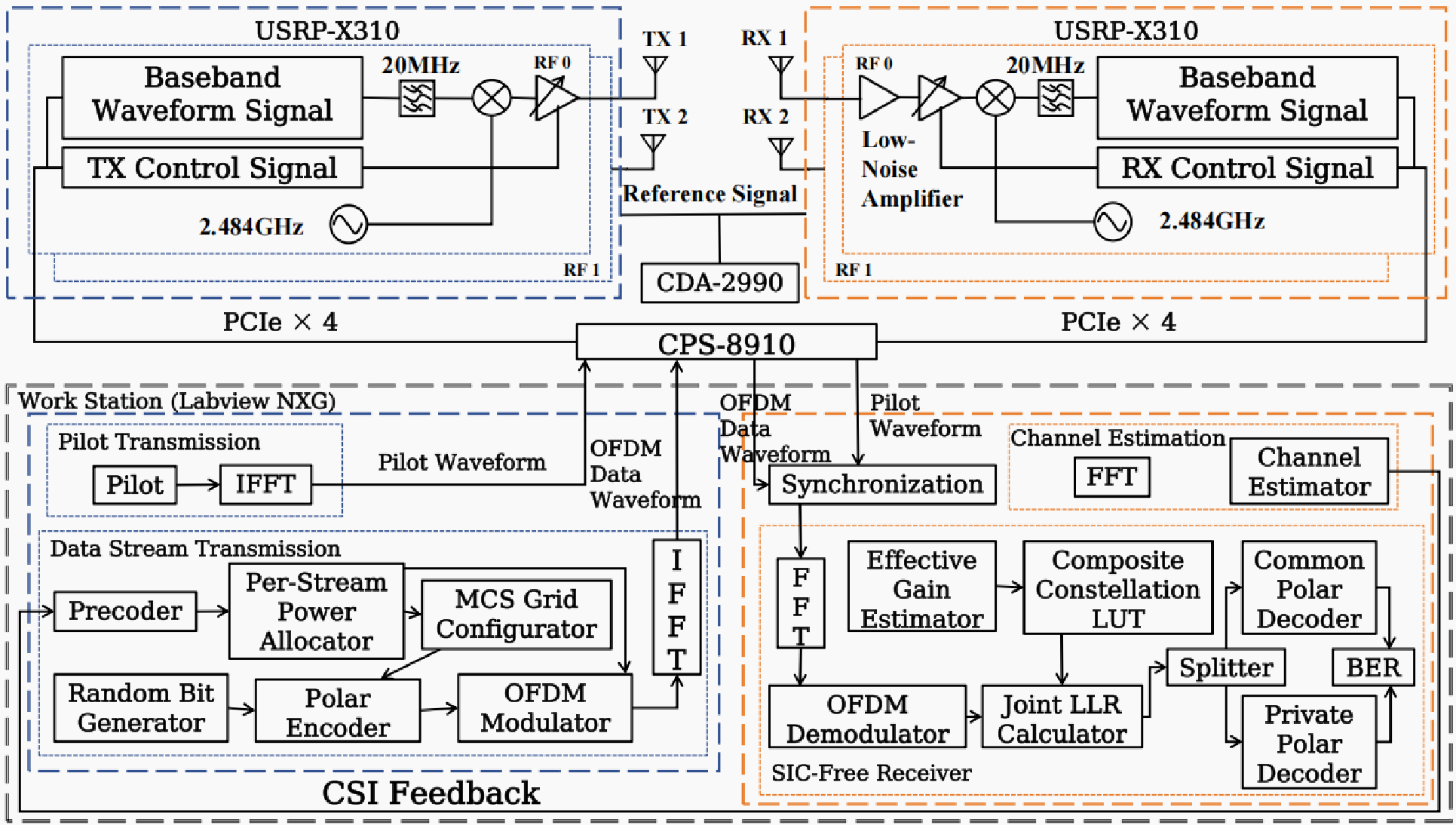} 
    \caption{SIC-free RSMA prototype block diagram.}
    \label{fig:Fig 2.1}
\end{figure}

Our prototype is built upon an SDR-based MU-MISO platform as shown in Fig. \ref{fig:Fig 2.1}. The transmitter uses one  National Instrument (NI) Universal Software
Radio Peripheral (USRP) driving two antennas to transmit the RSMA common and private streams toward the two receive antennas connected to the other NI USRP.

In the timing and I/O infrastructure part, a NI CDA-2990 clock distributor distributes a common 10 MHz reference signal to all USRPs for time synchronization. Then, a NI CPS-8910 PCIe switch box is applied to aggregate multiple PCIe \texttimes 4 links from the USRPs into PCIe \texttimes 8 to the workstation.

In the workstation part, the TX design follows the SIC-based RSMA prototype in \cite{10471302}. It generates random bit streams, applies polar encoding to form the common and private codewords, performs power allocation, and designs the precoder based on CSI feedback for user-1 and user-2. The resulting symbols are then go through OFDM modulation with Inverse Fast Fourier Transform and CP insertion. Meanwhile, pilots and a preamble are added for synchronization and channel estimation.

For transmission and throughput evaluation, the system is configured with an effective bandwidth $B_{eff}$ of 12 MHz and a subcarrier spacing of 312.5 kHz. Based on the data rate definition $R(m,r)$ from our system model, the corresponding operational data rates for the generated common and private codewords are summarized in the per-stream MCS grid in TABLE \ref{tab:mcs_levels}.

\begin{table}[!t]
  \caption{MCS grid implemented in the prototype.}
  \label{tab:mcs_levels}
  \centering
  \footnotesize
  \renewcommand{\arraystretch}{1.1}
  \setlength{\tabcolsep}{6pt}
  \begin{tabularx}{\columnwidth}{|X|X|X|X|}
    \hline
    \textbf{MCS Index} & \textbf{Modulation \(m\)} & \textbf{Code Rate \(r\)} & \textbf{Data Rate} \\
    \hline
    0 & BPSK & $1/2$ &  6 Mbps  \\
    \hline
    1 & BPSK & $3/4$ &  9 Mbps  \\
    \hline
    2 & QPSK & $1/2$ & 12 Mbps  \\
    \hline
    3 & QPSK & $3/4$ & 18 Mbps  \\
    \hline
    4 & 16QAM & $1/2$ & 24 Mbps  \\
    \hline
    5 & 16QAM & $3/4$ & 36 Mbps  \\
    \hline
  \end{tabularx}
\end{table}

Moving to the RX part from Fig. \ref{fig:Fig 2.1}, the receiver first performs frame synchronization to remove timing offsets and then applies a Fast Fourier Transform to obtain per-subcarrier symbols in the frequency domain. The upper part of the receiver illustrates the channel estimation procedure, where pilot symbols are fed into the channel estimator to obtain the estimated CSI, which is further used in the feedback loop for precoder design.

Utilizing the estimated subcarrier-dependent CSI, the receiver calculates the effective gains and constructs the composite constellation $\mathcal{S}_{k,n}$ as defined in the system model. Each baseband symbol on subcarrier $n$ is demapped as a symbol from this single composite alphabet. The demapper outputs a bit vector of length $b_c+b_p$ (with $b_c=\log_2 |\mathcal{X}_c|$ and $b_p=\log_2 |\mathcal{X}_{p,k}|$). A fixed bit order is used: the first $b_c$ bits are assigned to the common stream and the last $b_p$ bits are assigned to the private-$k$ stream. The resulting bits are separated into common and private bits. By collecting bits over all subcarriers, the bit-level codewords from the common and private streams are obtained and forwarded to the polar decoder.

As for the measurements using our RSMA prototype, we implement two RSMA receiver chains under the same frame structure. To align the theoretical expected throughput \eqref{eq:throughput} with our hardware measurements, we substitute the theoretical decoding success probabilities $\Pr(\cdot)$ with the empirical packet delivery ratios observed in the SDR. We evaluate both SIC and JD chains using this unified, measurement-aligned empirical sum-throughput metric:
\begin{equation}
\begin{aligned}
T_{\mathrm{mcs,emp}}^{\mathrm{JD}}\!\bigl(\mathbf{P},\mathcal{M}\bigr)
&= \frac{D_{s,c}^{\mathrm{JD}}}{T} \bigl(B_{eff} m_c r_c\bigr) \\
&\quad + \sum_{k=1}^{2} \frac{D_{s,k}^{\mathrm{JD}}}{T} \bigl(B_{eff} m_k r_k\bigr).
\end{aligned}
\label{eq_sum Tp eva}
\end{equation}
where $D_{s,c}^{\mathrm{JD}}$ denotes the number of runs in which the common stream is successfully decoded by both users,$D_{s,k}^{\mathrm{JD}}$ denotes the number of runs in which the private stream of user-k is successfully decoded, and $T$ indicates the total number of runs. This equation translates the hardware decoding outcomes into the continuous time domain (bps), enabling a direct and fair comparison of the SIC-based and SIC-free RSMA receivers.

\section{Measurement Campaign and Results}
\begin{figure}[!t] 
    \centering
    \includegraphics[width=1\columnwidth]{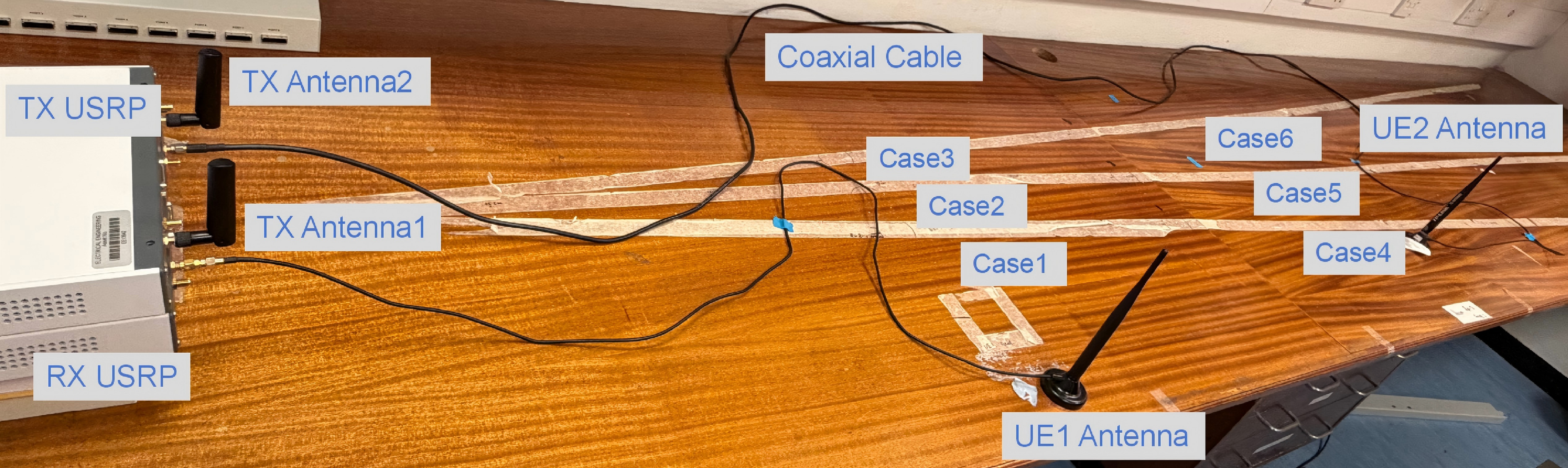} 
    \caption{RSMA measure campaign. User 1 is fixed, while user 2 is shown at Case 4 and can move from Case 1 to Case 6.}
    \label{fig:Measure camp}
\end{figure}

The experimental tests were settled on the lab bench as shown in Fig.~\ref{fig:Measure camp}. A full list of all parameters applied to our measurements is shown in TABLE \ref{tab:exp-params}.

\begin{table}[!t]
  \centering
  \caption{Parameters used in the experiments}
  \label{tab:exp-params}
  \setlength{\tabcolsep}{8pt}
  \renewcommand{\arraystretch}{1}
  \begin{tabular}{|p{0.35\linewidth}|p{0.2\linewidth}|p{0.21\linewidth}|}
    \hline
    \textbf{Parameter} & \textbf{Notation} & \textbf{Value} \\
    \hline
    Center frequency      & $f_c$ & 2.484\,GHz \\
    Transmit power        & $P_t$ & 16\,dBm \\
    TX antenna length     &       & 0.13\,m \\
    Fraunhofer distance   &       & 0.28\,m \\
    \hline
    Total bandwidth       &       & 20\,MHz \\
    \multirow[t]{4}{*}{Subcarriers}
      & Total $(N_c)$     & 64 \\
      & Data              & 48 \\
      & Pilot (FPS)       & 4 \\
      & Guard band        & 12 \\
    CP length             &       & 16 \\
    Effective bandwidth   & $B_{eff}$   & 12\,MHz \\
    OFDM symbols payload &      & 40 \\
    Experimental runs per case &   & 75\\
    \hline
  \end{tabular}
\end{table}

We then measure the average channel strength disparity $\alpha$ and spatial correlation $\rho$ from Case 1 to Case 6 shown in Fig. \ref{fig:Measure camp}. The relative channel strength disparity (in dB) between the two users is defined as $\alpha = 10 \log_{10} (\|\mathbf{h}_2\|^2 / \|\mathbf{h}_1\|^2)$, where $\mathbf{h}_1$ and $\mathbf{h}_2$ denote the estimated channel vectors for user 1 and user 2. To evaluate spatial correlation, we define $\rho = 1 - |\mathbf{h}_1^{\mathrm H} \mathbf{h}_2| / (\|\mathbf{h}_1\| \|\mathbf{h}_2\|)$. Under this definition, the channels become more aligned as $\rho$ approaches 0 and more orthogonal as it approaches 1. The empirical results averaged over all runs for each case are listed in TABLE \ref{tab:alpha-rho-cases}.

\begin{table}[!t]
  \centering
  \caption{Empirical averages of channel strength disparity $\alpha$ and spatial correlation $\rho$ for the cases in Fig.~\ref{fig:Measure camp}.}
  \label{tab:alpha-rho-cases}
  \setlength{\tabcolsep}{6pt}
  \renewcommand{\arraystretch}{1.15}
  \begin{tabular}{|c|c|c||c|c|c|}
    \hline
    \multicolumn{3}{|c||}{\textbf{Relative pathloss}} &
    \multicolumn{3}{c|}{\textbf{Spatial correlation}} \\
    \hline
     & \textbf{Cases} & $\boldsymbol{\alpha}$ [dB] &
     & \textbf{Cases} & $\boldsymbol{\rho}$ \\
    \hline
    \multirow{3}{*}{low}
       & 1 & $-1.34$  & \multirow{2}{*}{low} & 1 & 0.28 \\ \cline{2-3}\cline{5-6}
       & 2 & $-1.63$  &                       & 4 & 0.19 \\ \cline{2-3}\cline{4-6}
       & 3 & $-1.19$  & \multirow{2}{*}{mid} & 2 & 0.52 \\ \cline{1-1}\cline{2-3}\cline{5-6}
    \multirow{3}{*}{high}
       & 4 & $-11.74$ &                       & 5 & 0.49 \\ \cline{2-3}\cline{4-6}
       & 5 & $-11.87$ & \multirow{2}{*}{high}& 3 & 0.81 \\ \cline{2-3}\cline{5-6}
       & 6 & $-11.52$ &                       & 6 & 0.79 \\
    \hline
  \end{tabular}
\end{table}

\begin{figure}[!t] 
    \centering
    \includegraphics[width=0.9\columnwidth]{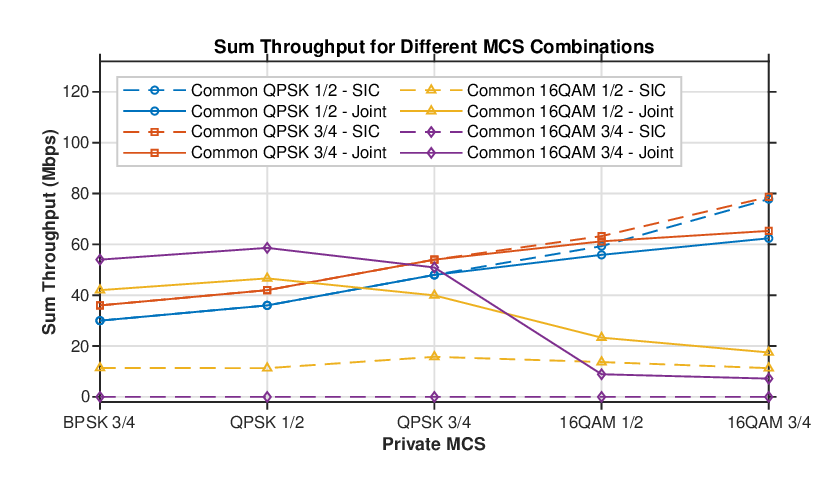} 
    \caption{Measured sum throughput versus private-stream MCS for Case 1 under SIC and JD.}
    \label{fig:Result1_1}
\end{figure}

\begin{figure}[!t] 
    \centering
    \includegraphics[width=0.9\columnwidth]{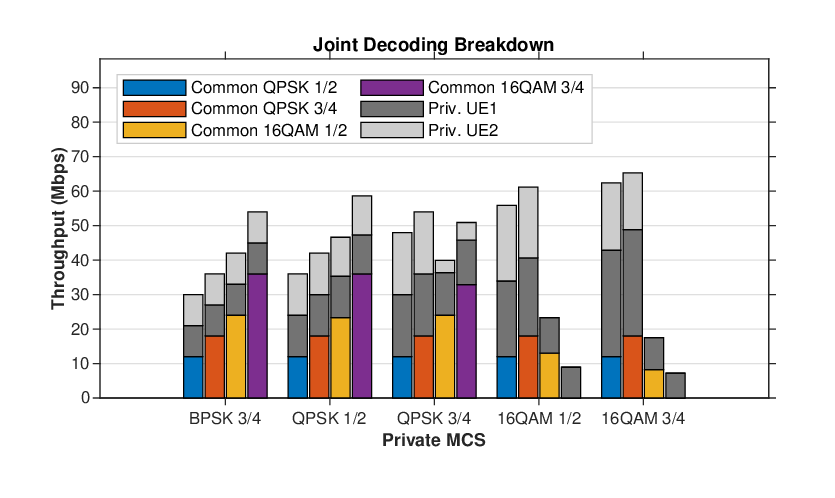} \\
    \includegraphics[width=0.9\columnwidth]{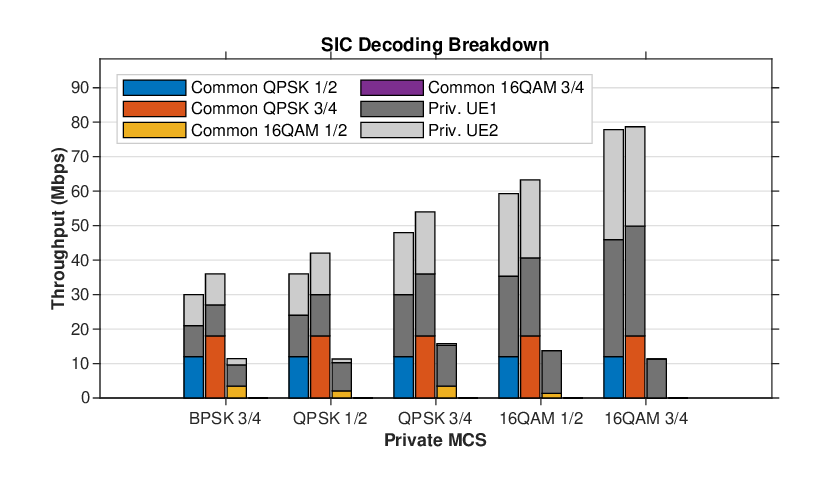} 
    \caption{Case 1 throughput decomposition: top, JD; bottom, SIC.}
    \label{fig:Result1_breakdown}
\end{figure}

We exhaustively search the MCS grid shown in TABLE \ref{tab:mcs_levels}. For each candidate MCS selection, decoding success probabilities are obtained empirically from measurements, and the resulting sum throughput is computed accordingly. The sum throughput performance for Case 1 as a function of the MCS levels is plotted in Fig. \ref{fig:Result1_1}, and throughput decomposition into common and two private contributions for two receivers are plotted in Fig. \ref{fig:Result1_breakdown}.

\begin{figure*}[!t]
    \centering
    \includegraphics[width=0.46\textwidth]{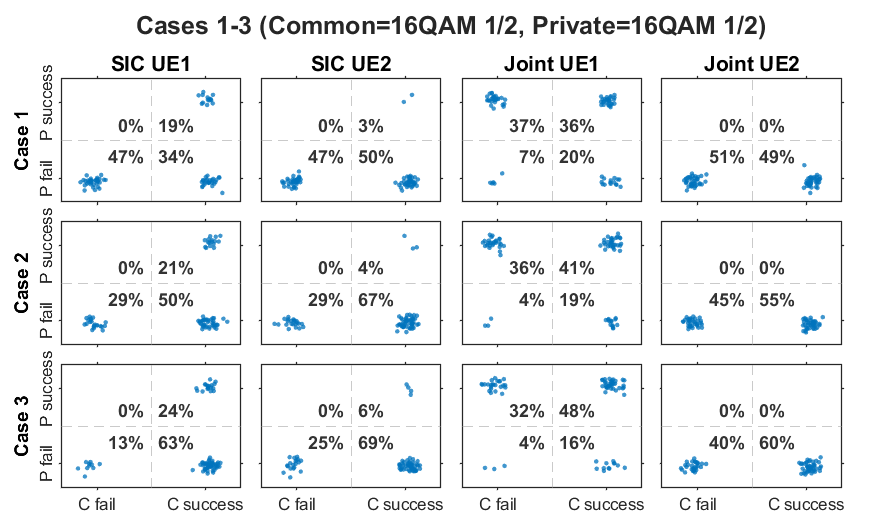}\hfill
    \includegraphics[width=0.46\textwidth]{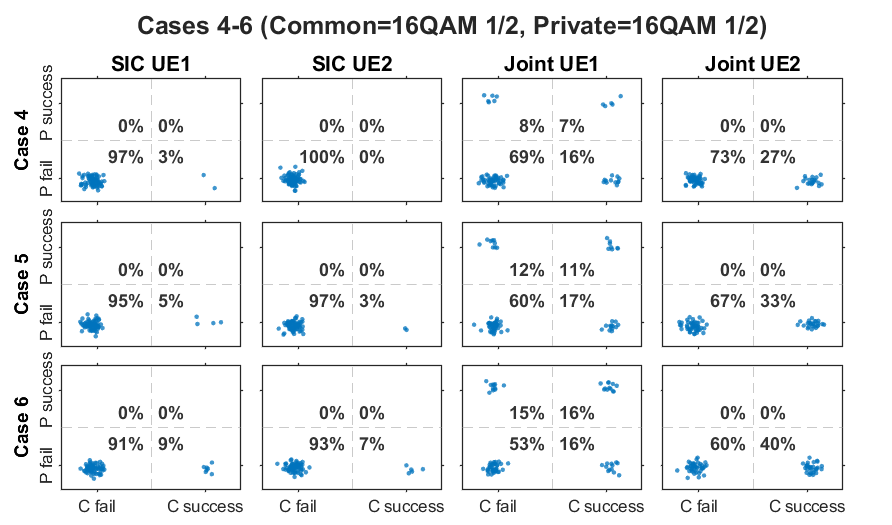}
    \caption{Four-quadrant decoding outcomes for SIC and JD across Cases 1--6. Left: Cases 1--3. Right: Cases 4--6.}
    \label{fig:Result2}
\end{figure*}

Fig. \ref{fig:Result1_1} shows that SIC achieves the highest peak throughput only when the common stream uses a moderate MCS, namely QPSK 3/4, whereas JD remains more robust as the common-stream MCS becomes more aggressive. When the common stream uses QPSK, the sum throughput generally increases with the private-stream MCS. The best measured operating point is obtained at common QPSK 3/4 and private 16QAM 3/4, where SIC achieves 78.69 Mbps and JD achieves 65.31 Mbps. As the common-stream MCS becomes more aggressive, however, the gap between SIC and JD is increasingly driven by reliability, i.e., successful common-message decoding, rather than peak throughput. With common 16QAM 1/2, SIC peaks around private QPSK 1/2 and then drops to 17.49 Mbps at private 16QAM 3/4. With common 16QAM 3/4, SIC collapses to nearly zero throughput across the private-MCS sweep. JD remains non-zero under common 16QAM 3/4, although it also degrades at high private orders because the composite constellation becomes very dense. The detailed decomposition in Fig. \ref{fig:Result1_breakdown} explain this trend. Under SIC, private-stream performance depends strongly on successful common decoding and accurate interference cancellation; hence, fragility in the common stream quickly removes both common and private contributions. In contrast, JD extracts decoding information directly from the composite constellation, allowing private-stream contributions to persist even when the common stream is near threshold.
To further demonstrate the superiority of JD over SIC in resisting error propagation, we present the four-quadrant analysis in Fig. \ref{fig:Result2} for all six cases in TABLE \ref{tab:alpha-rho-cases}. Each quadrant summarizes decoding success/fail outcomes for each user under a fixed MCS (common, private = 16QAM~1/2).

For the SIC-based receiver, the distributions move rapidly toward joint failure as channel difficulty increases. In the hardest case, the mass concentrates in the quadrant where both common and private fail for both users, often above 90\%. JD shows a different transition. In the easier cases, a large fraction remains in the quadrants with private success, and user 1 exhibits more than 30\% private-only success in Case~1 to~3. Even under worse channel conditions, JD avoids the near-total collapse seen with SIC and maintains private-message decodability when the common stream is stressed, effectively shifting outcomes from 'both-fail' to 'private-only success'.

Finally, we introduce coded BER versus SNR. The BER considered here is coded BER, as it directly determines the subsequent packet success rate and overall sum throughput.
\begin{figure}[!t] 
    \centering
    \includegraphics[width=0.8\columnwidth]{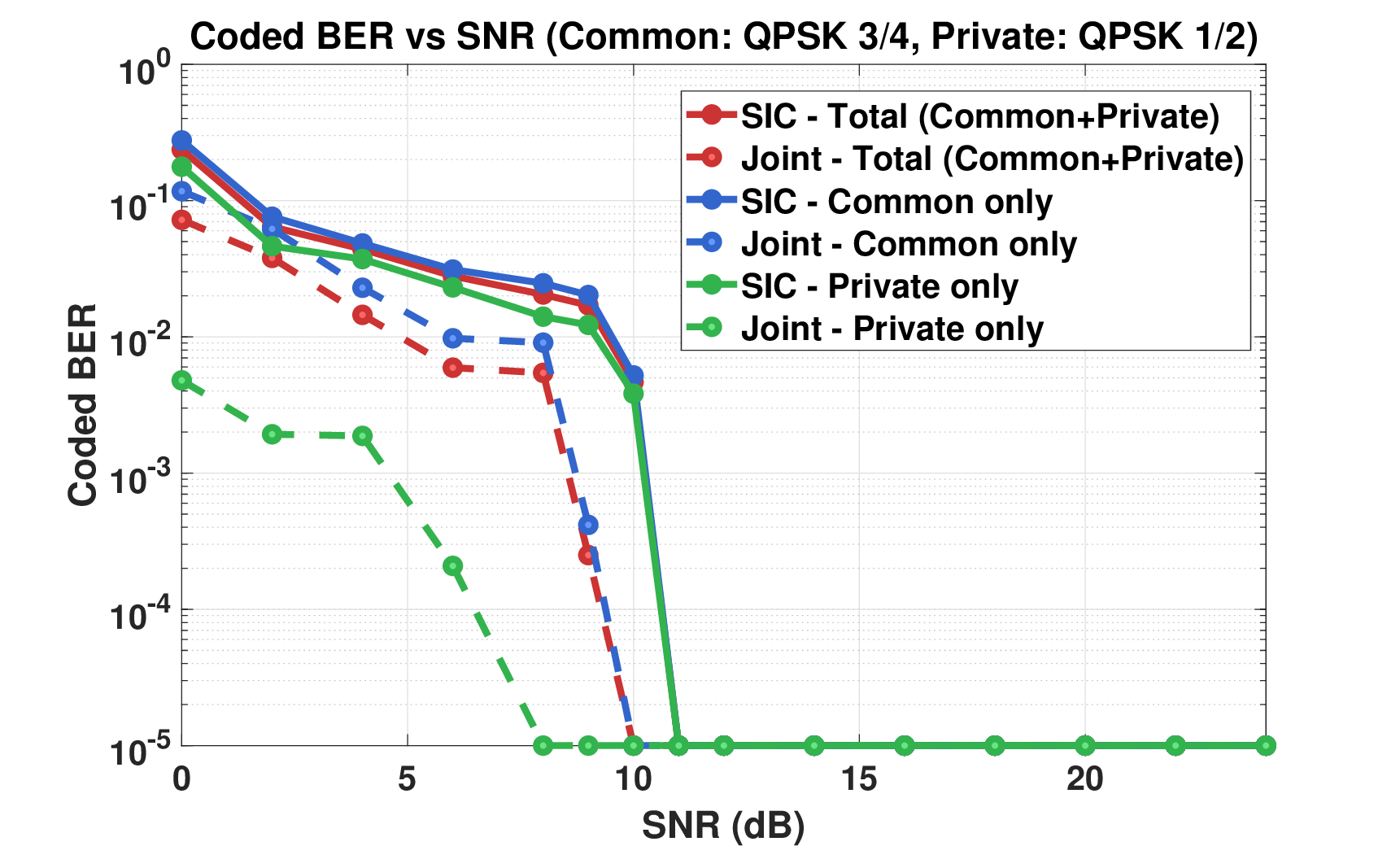} 
    \caption{Coded BER versus demapper-input SNR for SIC and JD.}
    \label{fig:Result3_1}
\end{figure}
The objective is to show the SNR advantage of JD over SIC under the same MCS group (Common QPSK 3/4, private QPSK 1/2). 
Fig. \ref{fig:Result3_1} compares coded BER versus the demapper input SNR $\gamma$, and we summarise the improvement using the threshold gain
$\Delta\gamma(\beta) = \gamma_{\mathrm{SIC}}(\beta) - \gamma_{\mathrm{JD}}(\beta),$
which is the SNR reduction required by JD to reach the same coded BER target $\beta$. For the common stream, JD provides a modest gain, about 1.2~dB at $10^{-3}$ and about 1~dB at $10^{-5}$. For the private streams, the gain is much larger, about 5.2~dB at $10^{-3}$ and about 3~dB at $10^{-5}$, which is consistent with SIC suffering error propagation from the common stage. When both common and private streams must meet the target simultaneously, the overall gain is about 1.7~dB at $10^{-3}$ and about 1~dB at $10^{-5}$, mainly limited by the common-stream threshold. Overall, the $\Delta\gamma$ results demonstrate that JD strengthens private-stream robustness and prevents the sharp failures that occur under serial SIC when the common stream is stressed.

\section{Conclusion}
In this paper, we implemented an SDR-based SIC-free RSMA receiver prototype based on joint demapping. Although JD incurs higher computational complexity than serial SIC due to joint bit-vector evaluation, our measurements demonstrate its practical feasibility and improved reliability across a wide range of MCS selections. By mitigating the error propagation inherent in SIC, the JD-based receiver provides a receiver-side enhancement that delivers smoother sum-throughput behavior and reduced decoding thresholds without requiring transmitter-side modifications.

\balance
{\bibliographystyle{IEEEtran}
\bibliography{Reference/Reference}
}

\end{document}